\documentclass[aps,prb,twocolumn,superscriptaddress,showpacs
]{revtex4-2}	%
\usepackage[ngerman,english]{babel} %
\usepackage[utf8]{inputenc}

\usepackage{amsmath,amssymb} %

\usepackage[pdfstartview=FitH,      %
            breaklinks=true,        %
            bookmarksopen=true,     %
            bookmarksnumbered=true  %
            ]{hyperref}				%

\let\oldcite\cite
\renewcommand*\cite[1]{\textsuperscript{\oldcite{#1}}}

\usepackage{wrapfig}
\usepackage{floatrow}    %
\usepackage{subfig} 

\usepackage{mwe}
\usepackage{graphicx}

\usepackage{array}
\usepackage{multirow}

\usepackage{adjustbox} 		%
\usepackage{upgreek}		%
\usepackage{units}		%
\usepackage{mathtools}			%
\usepackage{xargs}                      %

\usepackage[colorinlistoftodos,prependcaption,textsize=tiny]{todonotes}
\newcommandx{\unsure}[2][1=]{\todo[linecolor=red,backgroundcolor=red!25,bordercolor=red,#1]{#2}}
\newcommandx{\change}[2][1=]{\todo[linecolor=blue,backgroundcolor=blue!25,bordercolor=blue,#1]{#2}}
\newcommandx{\info}[2][1=]{\todo[linecolor=OliveGreen,backgroundcolor=OliveGreen!25,bordercolor=OliveGreen,#1]{#2}}
\newcommandx{\improvement}[2][1=]{\todo[linecolor=Plum,backgroundcolor=Plum!25,bordercolor=Plum,#1]{#2}}
\newcommandx{\thiswillnotshow}[2][1=]{\todo[disable,#1]{#2}}

\usepackage{lipsum}
\usepackage{booktabs}  %

\usepackage{placeins}  %

\makeatletter
\def\p@paragraph{\thesection\,\thesubsection\,}%
\makeatother

 \makeatletter
\newcommand{\phantomlabel}[2]{
    \protected@write\@auxout{}{
        \string\newlabel{#2}{
            {\@currentlabel#1}{\thepage}
            {\@currentlabel#1}{#2}{}
        }
    }
    \hypertarget{#2}{}
}
\makeatother

\begin{document}

\title{Non-hermiticity in spintronics: oscillation death in coupled spintronic nano-oscillators through emerging exceptional points}

\author{Steffen Wittrock}%
\email{steffen.wittrock@helmholtz-berlin.de}\affiliation{Unit\'{e} Mixte de Physique CNRS, Thales, Universit\'{e} Paris-Saclay, 1 Avenue Augustin Fresnel, 91767, Palaiseau, France}\affiliation{Helmholtz-Zentrum Berlin für Materialien und Energie GmbH, Hahn-Meitner-Platz 1, 14109 Berlin, Germany}
\author{Salvatore Perna}\affiliation{Department of Electrical Engineering and ICT, University of Naples Federico II, 80125 Naples, Italy} 
\author{Romain Lebrun}\affiliation{Unit\'{e} Mixte de Physique CNRS, Thales, Universit\'{e} Paris-Saclay, 1 Avenue Augustin Fresnel, 91767, Palaiseau, France}
\author{Katia Ho}\affiliation{Unit\'{e} Mixte de Physique CNRS, Thales, Universit\'{e} Paris-Saclay, 1 Avenue Augustin Fresnel, 91767, Palaiseau, France}
\author{Roberta Dutra}\affiliation{Centro Brasileiro de Pesquisas Fésicas (CBPF), Rua Dr. Xavier Sigaud 150, Rio de Janeiro 22290-180, Brazil}
\author{Ricardo Ferreira}\affiliation{International Iberian Nanotechnology Laboratory (INL), 471531 Braga, Portugal }
\author{Paolo Bortolotti}\affiliation{Unit\'{e} Mixte de Physique CNRS, Thales, Universit\'{e} Paris-Saclay, 1 Avenue Augustin Fresnel, 91767, Palaiseau, France}
\author{Claudio Serpico}\affiliation{Department of Electrical Engineering and ICT, University of Naples Federico II, 80125 Naples, Italy} 
 \author{Vincent Cros}\email[]{vincent.cros@cnrs-thales.fr}\affiliation{Unit\'{e} Mixte de Physique CNRS, Thales, Universit\'{e} Paris-Saclay, 1 Avenue Augustin Fresnel, 91767, Palaiseau, France}

\date{February 1, 2023}

\definecolor{light-gray}{rgb}{0.96,0.96,0.96}

\begin{abstract}

\bfseries 

The emergence of exceptional points (EPs) in the parameter space of a non-hermitian (2D) eigenvalue problem is studied in a general sense in mathematical physics, and has in the last decade successively reached the scope of experiments. 
In coupled systems, it gives rise to unique physical phenomena, which  enable  novel approaches for the development of seminal types of highly sensitive sensors. 
Here, we demonstrate at room temperature the emergence of EPs in coupled spintronic nanoscale oscillators and hence exploit the system's non-hermiticity. 
We describe the observation of amplitude death of self-oscillations and other complex dynamics, and develop a linearized non-hermitian model of the coupled spintronic system, which properly describes the main experimental features. 
Interestingly, these spintronic nanoscale oscillators are deployment-ready in different applicational technologies, such as field, current or rotation sensors, radiofrequeny and wireless devices and, more recently, novel neuromorphic hardware solutions. Their unique and versatile properties, notably their large nonlinear behavior, open up unprecedented perspectives in experiments as well as in theory on the physics of exceptional points.  
Furthermore, the exploitation of EPs in spintronics devises a new paradigm for ultrasensitive nanoscale sensors and the implementation of complex dynamics in the framework of non-conventional computing.

\end{abstract}


\maketitle

{\em Exceptional points} (EPs) are singularities in the parameter space of a system corresponding to the coalescence of two or more eigenvalues and the associated eigenvectors \cite{Kato1982,Heiss2000,Heiss2001,Heiss2012,Dolfo2018,Kawabata2019}.  
They are a peculiar feature of nonconservative (open) systems that have both loss and gain and they emerge when these two effects compensate.  
From the fundamental point of view, EPs play an important role in the area of non-Hermitian quantum theory based on  $\mathcal{PT}$-symmetric Hamiltonians (with simultaneous parity-time invariance) \cite{Bender2019}. 
In this context, they occur at the phase transitions between broken-unbroken $\mathcal{PT}$-symmetry. 
While initially EPs were regarded as a mathematical-physics concept, in the last decade there has been a growing interest
on EPs from the experimental point of view in such areas as atomic spectra measurements \cite{Cartarius2007}, microwave cavity experiments \cite{Dembowski2001,Dembowski2003,Chen2017,Doppler2016,Hodaei2017},  chaotic optical microcavities \cite{Lee2009} or optomechanical systems \cite{Xu2016}. 
Interestingly, exceptional points arise also in classical systems, such as coupled electric oscillators \cite{Heiss2004,Stehmann2004}, 
 optical systems \cite{Rueter2010}, classical spin
dynamics \cite{Tserkovnyak2020, Galda2019}, and general dissipative classical systems \cite{Ryu2015}. 

From the applications point of view, the strong sensitiveness of eigenvalues to perturbations near EPs has been used to devise new types of sensors with unprecedented sensitivity \cite{Chen2017,Wiersig2014,Wiersig2020}. 
This was demonstrated in highly sensitive optical nanoparticle detection \cite{Zhang2015,Wiersig2016}, in laser gyroscopes \cite{Ren2017}, in optically pumped semiconductor rings for temperature detection\cite{Hodaei2017}, and in coupled microcantilevers for ultrasensitive mass sensing\cite{Spletzer2006}, however non of them being adapted to CMOS compatible systems. 

\begin{figure}[bth!]
    \centering
    \includegraphics[width=0.7\columnwidth]{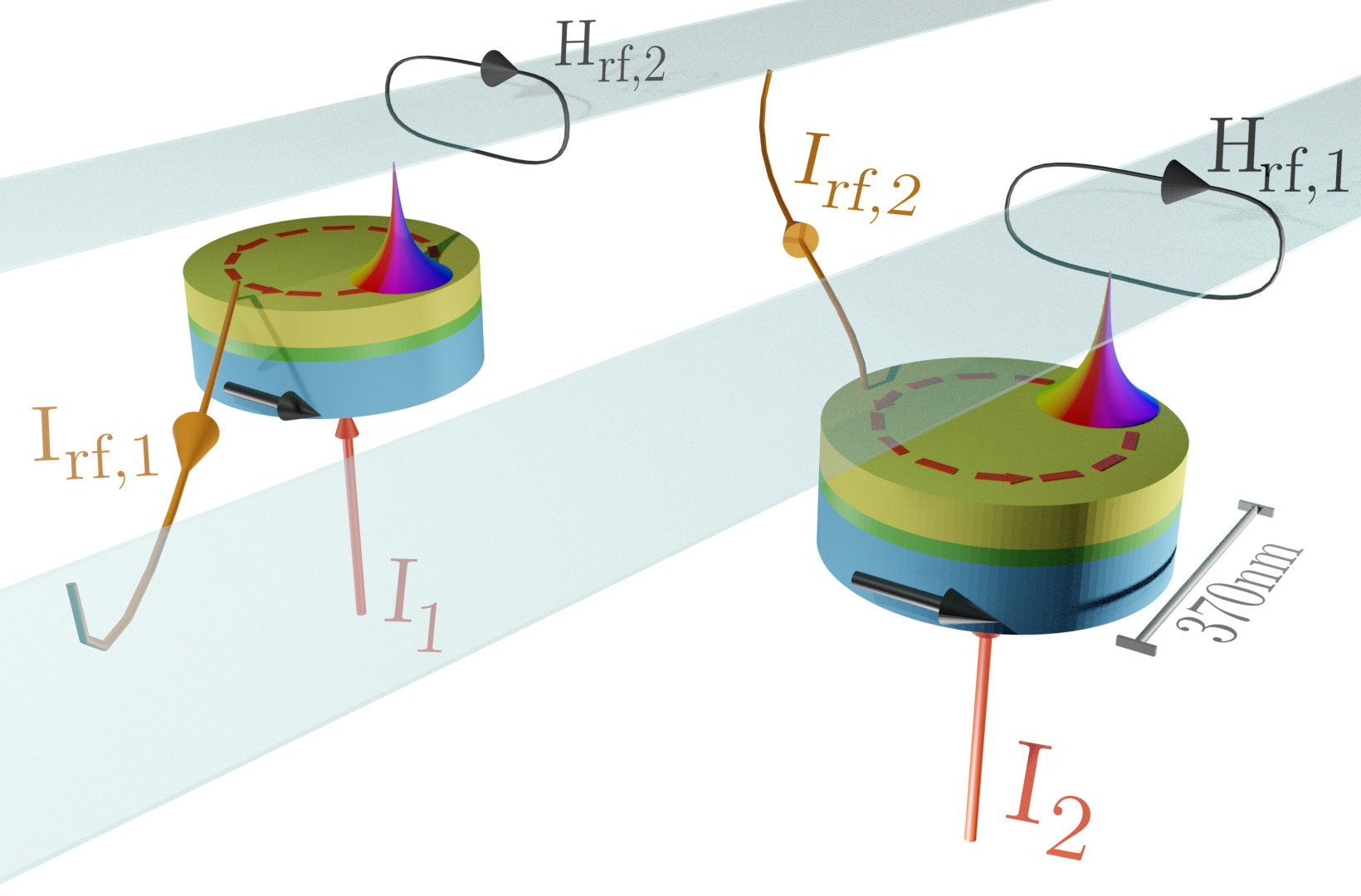}
    \caption[]{ Schematic view of the spintronic EP system made of coupled spintronic nano oscillators.  The coupling, which is designed to be symmetric, is obtained by feeding strip-line antennas above each oscillator with the microwave current $I_{\text{rf}}$ generated by the spin torque dynamics occurring in the other oscillator. This generates a magnetic rf field, $H_{\text{rf}}$, that hence generates the coupling \cite{Singh2019}. The coupling is nonlocal and can be performed over long distances.  }
    \label{fig:circuit}
\end{figure}

In this study, we demonstrate how the presence of EPs can be used to control the oscillating state in a system of two coupled spin-torque nano-oscillators (STNOs) (see fig. \ref{fig:circuit}). 
These are typical, CMOS-compatible, spintronic nanoscale devices in which the magnetization dynamics in a thin layer can be converted into  electrical microwave signals\cite{Locatelli2013}. They have both loss, associated with the damping of magnetization oscillations, and gain, provided by the transfer of spin angular momentum (through the spin torque effect) from a spin polarized current injected into the device (see Methods section for further details). 
In this respect, coupled STNOs are archetypal non-hermitian systems to evidence the relevance of EPs in spintronics. While maintaining the generality of the results, we use in practice STNOs based on the spin torque gyrotropic dynamics of a magnetic vortex core \cite{Locatelli2013} (later on labelled as STVOs), since they are today the ones showing the best rf performance\cite{Tsunegi2014}. %

In the last couple of years, there has been growing attention on EPs and the effects of non-hermiticity in the area of magnetism. 
Different approaches have been used exploiting %
the coupling of magnons to other distinct quantum systems, such as phonons \cite{Berk2019} or explicitly  photons\cite{Harder2017,Zhang2017,Yang2020,Zhang2019}, where great advance could be achieved in the field of cavity-magnonics\cite{Rameshti2021}. 
In spintronics, the interest on EPs started with theoretical works on classical spin dynamics\cite{Lee2015,Galda2016,Galda2019, Tserkovnyak2020, Yu2020,Yu2020,Proskurin2021} and lately also includes spin wave physics\cite{Wang2020,Wang2021}. 
The potential of non-hermitian effects in coupled STNOs has been theoretically emphasized in larger dimensional arrays, in which nontrivial non-hermitian topological phases can emerge\cite{Flebus2020,Gunnink2022}, an aspect that is yet to be explored experimentally. 
It is to be noted that all works predominantly focus on local coupling mechanisms. 
Only recently, also based on a local coupling, the emergence of an EP could be demonstrated experimentally in magnonic $\mathcal{PT}$-symmetry devices\cite{Liu2019}. 
However, the approach stays passive, i.e. the magnetic dissipation (damping) in two coupled magnetic layers is fixed and the system control (through the coupling) was realized through the choice of the thickness of the separating layer between the  ferromagnetic thin films. 
The experimental study of EPs in coupled discrete nano-devices (with gain) has been so far overlooked and an experimental control of the coupled dynamics on a nonlocal circuit level is completely missing. 
The purpose of our study is to experimentally demonstrate that, by tuning the dc currents injected into the two STNOs, the position of an EP can be finely controlled leading to the phenomenon of {\em amplitude death}\cite{Aronson1990}. 
This describes the vanishing of the oscillation amplitudes of the coupled STNOs, despite the increase of the spin torque (gain) in one oscillator. %
As shown later, such amplitude death occurs for a certain specific range of injected dc current values. 

We believe that the presented results create an important connection between the non-hermitian physics of EPs and spintronics, an area of research that has crucial technological implications for data storage and processing\cite{Hirohata2020,Chumak2017}, sensor technology\cite{Hirohata2020,Garcia2021_noise-reduction_APL}, wide-band high-frequency communications\cite{Choi2014,Jenkins2016,Ruiz-Calaforra2017,Kreissig2017-IEEE,Louis2017,Litvinenko2019}, and, 
more recently, bio-inspired networks for neuromorphic computing beyond CMOS \cite{Torrejon2017,Romera2018}. %
Up to now, coupled STNOs have been studied mainly with respect to the phenomenon of mutual synchronization \cite{Kaka2005,Mancoff2005,Slavin2009,Lebrun2017,Tsunegi2018} and non-hermitian aspects are neglected or are about to be theoretically discovered\cite{Flebus2020,Gunnink2022}. 
Our observation describes, up to our knowledge, the first experimental evidence to exploit EPs in coupled spintronics nano-devices.

 \vspace{2ex}
\paragraph*{\bf Theoretical modelling of EPs in spintronics}~ 

We first  theoretically study the regime of small amplitude oscillations of the  two coupled STVOs around their rest positions. From the linearized equations governing these small oscillations, the 
condition for an EP to emerge can be determined, leading to a formula that connects,
at the EP, the relevant parameters of the coupled oscillators: frequencies, 
gain/loss parameters, and coupling coefficient. The position of an EP in the parameter space can hence be controlled in order
to determine the interval of injected current values in which 
amplitude death occurs. 
The gain mechanism, counteracting the natural dissipation and enabling self-sustained oscillations in each STVO,  
is provided by the spin-transfer torque that is proportional to the injected dc current. Self-oscillations set in %
when the injected current $I$ is larger than a critical (threshold) current $I_c$, which corresponds to 
the exact compensation of gain and loss\cite{Slavin2009}.  

Gyrations of the vortex core around the symmetry axis of each oscillator are modeled by the Thiele-like theory for which the overall state of the oscillators is given by the in-plane  displacements  $(\boldsymbol{\rho}_{1}, \boldsymbol{\rho}_{2})$ of the vortex cores from the center of the corresponding devices (for details see Methods \ref{sec_methods:theory}). %
The coupling between the two STVOs, which is assumed to be symmetric, is obtained by feeding 
strip-line antennas above each oscillator with the microwave voltage generated by the other, that in turn gives rise to a rf magnetic field
(see fig. \ref{fig:circuit}). For vortex core displacements sufficiently small compared to the device radius, it is reasonable to assume a linear coupling\cite{Aronson1990,Balanov2008-synchronization} 
between the STVOs, reflecting the relevant range of the performed experiments.  %
Importantly, the coupling has both dissipative and conservative terms that are described by the coefficients $k_d$ and $k_c$, respectively.

The linearized Thiele equation governing the regime of small oscillations of the vortex cores around the rest position $\boldsymbol{\rho}_{1,2}=0$, written
in terms of the complex state variables 
$z_l = x_l+ iy_l$ (with $l=1,2$) associated to the $l$-th vortex core $x$- and $y$-axis position, reads
\begin{align}
\frac{d}{dt} \begin{bmatrix} z_1 \\ z_2 \end{bmatrix}   = i A \cdot \begin{bmatrix} z_1 \\ z_2 \end{bmatrix} \, ,
\label{eq:AdynSyn}
\end{align}
\hspace{0.25cm} with
\begin{align}
A = \begin{bmatrix}  \omega_1 -i \beta_1  &   k   \\
k  &  \omega_2 -i\beta_2    \end{bmatrix}
\label{eq:eps_matrix}  \, ,
\end{align}
where $k=k_c -i k_d $, $\omega_l$ are the angular frequencies of vortex free oscillations, and $\beta_l$ are the loss/gain parameters. These latter parameters are given by $\beta_{l} = C_l I_l - d_l\omega_l$, where
$d_l$ are the damping constants and $C_l$ are parameters determining the efficiency of the 
spin torque effect.

The matrix $A$ in eq. \eqref{eq:eps_matrix} is non-hermitian and has indeed the  typical form for systems exhibiting EPs \cite{Kato1982,Heiss2012}. 
In order to study the natural frequencies of the system \eqref{eq:AdynSyn}, we assume 
a dependence of $z_{1,2}$ on the time of the type $e^{i\nu t}$. The natural frequencies $\nu_{1,2}$  are then obtained as eigenvalues of the matrix $A$ and they
are given by the following formula: 
\begin{align}
\nu_{1,2} = \bar{\omega} - i\bar{\beta} \pm \sqrt{ k^2 + \left(  \tilde{\omega} -i\tilde{\beta} \right)^2} \,,
\label{eq:natural_frequencies}
\end{align}
where
\begin{align*}
\bar{\beta} = {(\beta_1+\beta_2)}/{2} \, , \,\,\, \bar{\omega} = {(\omega_1+\omega_2)}/{2} \, , \\
\tilde{\beta} = {(\beta_1-\beta_2)}/{2} \, , \,\,\, \tilde{\omega} = {(\omega_1-\omega_2)}/{2} \, .
\end{align*}
Stability of solutions is given by the following condition: 
\begin{align}
\text{Im}(\nu) \geq 0 \,\, \Longleftrightarrow \,\, \text{Re}(i \nu) \leq 0 \, . \label{eq:stability-criterion}
\end{align}

By definition, EPs emerge when two natural frequencies coalesce along with the corresponding 
eigenvectors. This occurs when the square root term in eq. \eqref{eq:natural_frequencies} is zero, leading to the following condition on the parameters to obtain an EP:
\begin{align}
k_c - i k_d  = \pm \left[ \frac{(\beta_1-\beta_2)}{2}  + i\frac{(\omega_1-\omega_2)}{2}  \right] \, ,
\label{eq:eps_condition}
\end{align}
where we have explicitly expressed  $k$, $\tilde{\omega}$ and $\tilde{\beta}$.

\begin{figure}[bth!]
    \centering
    \subfloat[ \label{fig_exceptional:theory-1}]
{%
\includegraphics[width=0.48\textwidth]{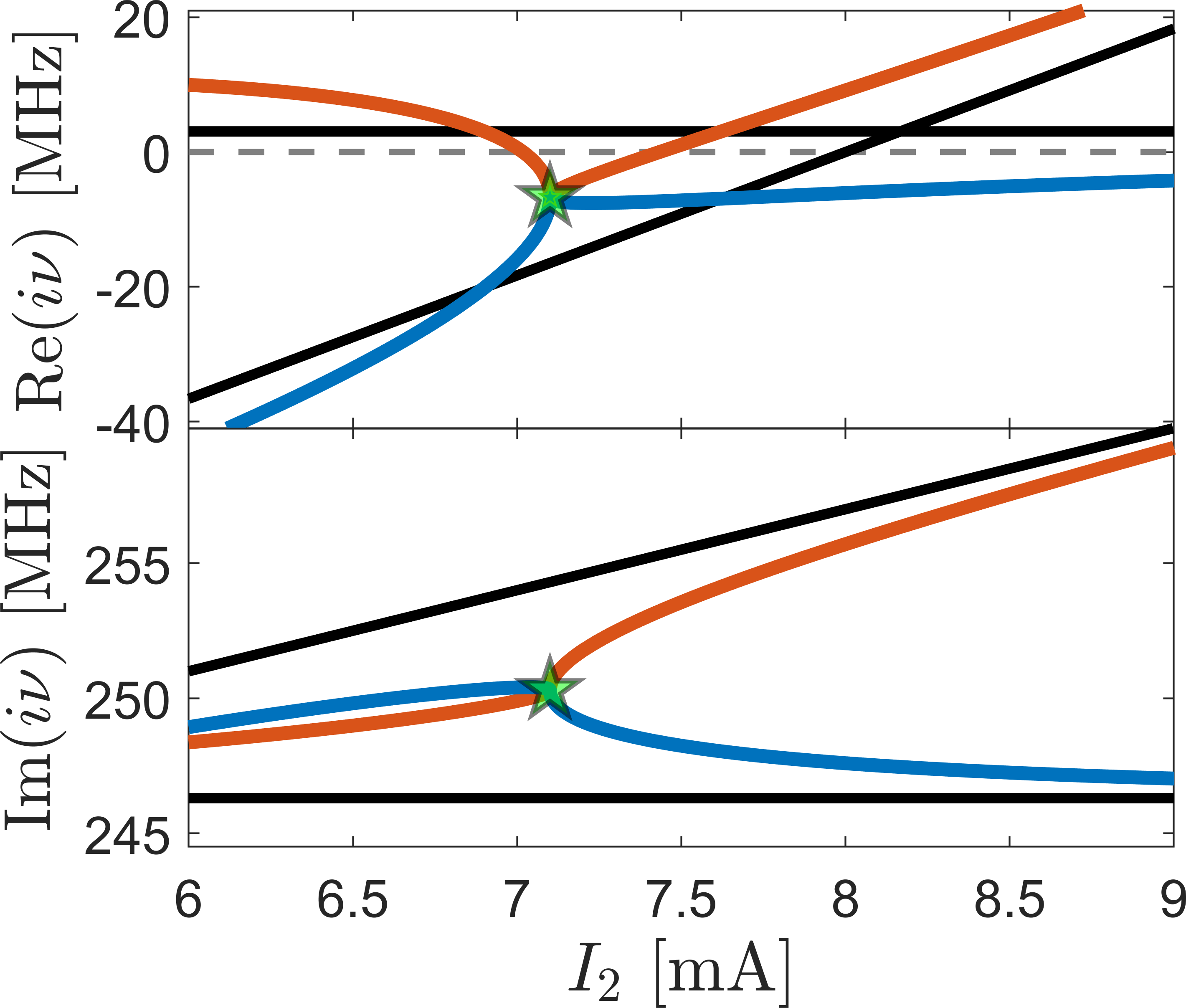}  }
  \subfloat[ \label{fig_exceptional:theory-2}] 
 { %
 \includegraphics[width=.48\textwidth]{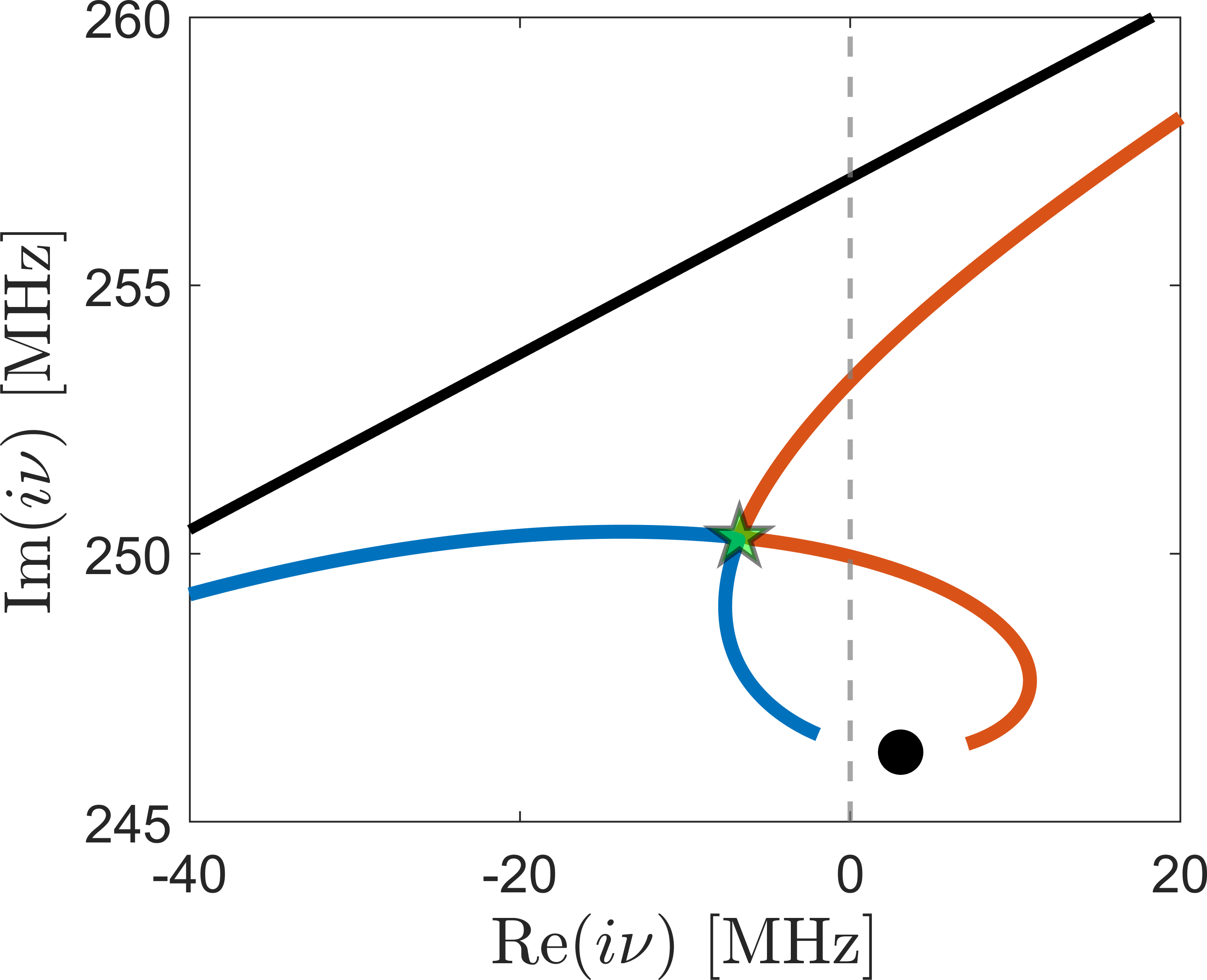}  } 
    \caption[]{Eigenvalues $i\nu_{1,2}$ from eq. \eqref{eq:natural_frequencies} when the EP is placed at $I_1^*=I_{2,\text{EP}}=7.1$ mA (green star). (a) Real and imaginary part of the eigenvalues as a function of the dc current $I_2$. (b) Eigenvalues in the complex plane (Re$(i\nu)$, Im$(i\nu)$). Black lines in (a) as well as black line and symbol in (b) refer to eigenvalues computed in the uncoupled case.
    The system parameters can be found in the Methods section \ref{sec_methods:theory}. The resulting coupling constant is $k_{EP} = 9.76 - 25.13i$.}
    \label{fig:EP1}
\end{figure}

We consider here the case that will be also presented in the experimental section: 
 the current $I_1$ is fixed to a value $I_1^*$  and 
the second current $I_2$ is swept from values below to values above the threshold current  $I_{c,2}$. 
By using the condition \eqref{eq:eps_condition}, the values of the parameters can be adjusted to have an EP at the desired value of the current $I_2$. 
The effect of the EP on the  eigenvalues of the matrix $A$ is illustrated 
in fig. \ref{fig:EP1}.  The black curves in fig. \ref{fig_exceptional:theory-1} are the real and imaginary parts of 
the STVOs' eigenvalues when there is no coupling ($k_c=k_d=0$). When coupling 
is taken into account, an EP exists at $I_{2,\text{EP}}=7.1\,$mA (green star), and it has the 
effect of attracting the two eigenvalues to one point in the complex plane. 
According to eq. \eqref{eq:natural_frequencies}, if, at the EP, $\bar{\beta}= (\beta_1+\beta_2)/2$ is
negative (this occurs when $\beta_2 <0$, $|\beta_2| > \beta_1 >0$),
then both eigenvalues $i\nu$ have negative real parts. This is also visible in fig. \ref{fig_exceptional:theory-2} where the eigenvalues are plotted in the complex plane (Re$(i\nu)$, Im$(i\nu)$): In the proximity of the EP, both eigenvalues are in the plane Re$(i\nu)<0$. 
This implies that the rest position of the two oscillators is stable, 
leading to the disappearance of both STVOs' oscillations. 
This phenomenon is called \textit{amplitude death} and, as we have illustrated above, it can be controlled by the appropriate placing of the EP in the parameter space. 

Since the condition for the onset of an EP is very sensitive to perturbations, it might happen in experiments that the EP is not  reached in a strict sense, as we discuss hereafter in the analysis of our experimental data. 
Nevertheless, if the parameters are such that the condition \eqref{eq:eps_condition} 
is nearly verified, the amplitude death phenomenon is expected to be reliably observed as well. 

It is important to note that, when the rest position is stable in terms of the linearized model, we find that this stability is also exhibited by the rest position in the full nonlinear equations (see Methods \ref{sec_methods:nonlinear_simulations}). An important consequence is that, for predicting the phenomenon of amplitude death, the linear theory is strictly appropriate. On the other hand, when the real part of the eigenvalue $i\nu$ becomes positive -- this  happens when Re$(i\nu)$ crosses zero in fig. \ref{fig:EP1}  -- the rest state becomes unstable. The regime that sets in after instability has an amplitude determined by the nonlinear saturation term in the Thiele equation and an approximate frequency of Im$(i\nu)$ at the aforementioned crossing. This phenomenon is referred to as a Hopf bifurcation\cite{Wiggins2003}. For values of parameters which correspond to a Hopf bifurcation point and no other bifurcations take place, the linear analysis can be  used to estimate the frequency of the self-oscillating regimes by considering the imaginary parts of the natural frequency $i\nu$ at the Hopf bifurcation. In the assessment of the experimental results, this concept is applied in order to identify the appropriate parameter values describing the amplitude death region as a function of the injected currents.

\vspace{2ex}
\paragraph*{\bf Experimental emergence of EPs}~

After having theoretically established the condition for the existence of an EP, we describe the experimental results that  demonstrate the emergence of EPs and the correlated amplitude death regions in our coupled STVO system. 
All measurements have been conducted at room temperature. 
In the performed experiments,  the current injected into the STVO 1 is kept constant to $I_1^*$, %
 while sweeping the current $I_{2}$ injected into STVO 2. 
Note that the onset for self-sustained oscillations in the uncoupled case is $I_{c,1}\approx 6.95\,$mA and $I_{c,2}\approx 8\,$mA for STVO 1 and 2, respectively. 
The frequency evolution of the uncoupled STVOs with the applied current is similar (see Methods \ref{sec_methods:experiment}).

\begin{figure}[bth!]
  \centering  %
\subfloat[\label{fig:2nd_vsSTO2_8.0mA_sweepSTO2}]
{%
\includegraphics[width=0.51\textwidth]{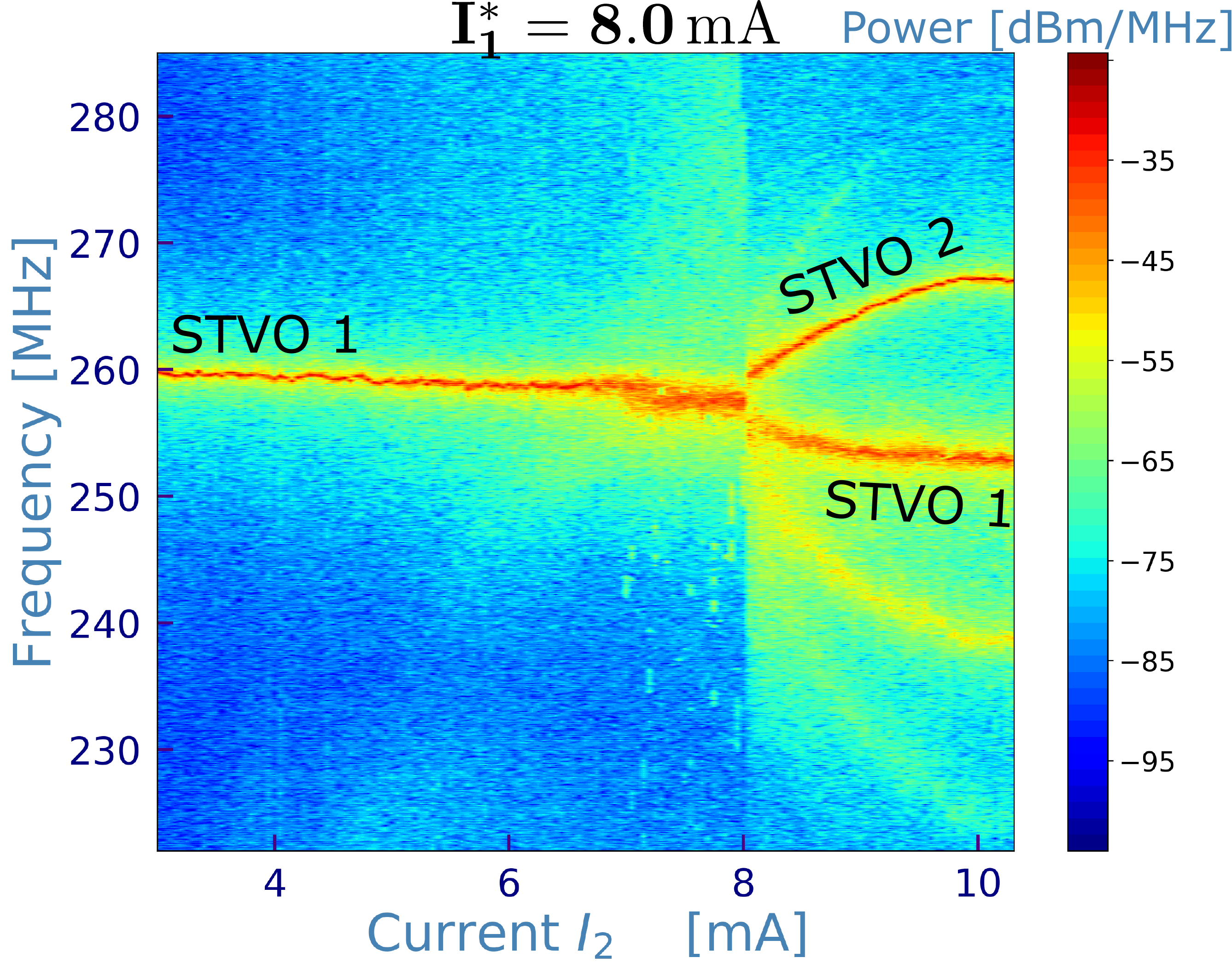} }
\subfloat[ \label{fig_exceptional:theory_8mA}]
{%
\includegraphics[width=.465\textwidth]{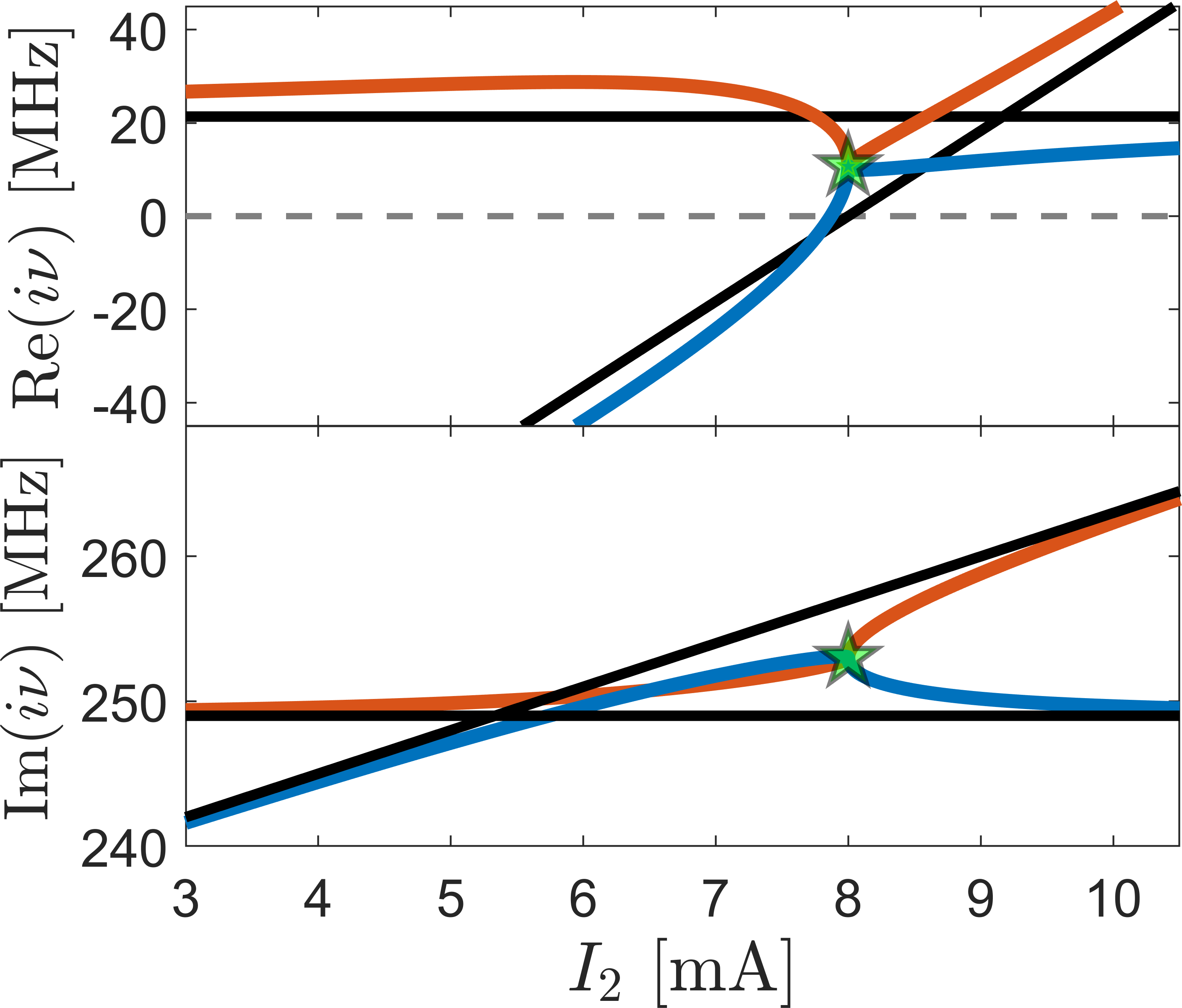}}
\caption[]{ Measured frequency spectra of the coupled system vs. current $I_{2}$ for $I_{1}^*=8\,$mA (a). The labeling corresponds to the STVO in which the oscillation mode is mainly localized.  (b) Corresponding theoretically determined evolution of eigenvalues. System parameters in the Methods section. The resulting coupling constant: $k_{EP} = 10.68 - 25.13i$. 
}      %
  \label{fig_exceptional:2nd_vsSTO2_sweepSTO2_8mA}
\end{figure}

In fig. \ref{fig:2nd_vsSTO2_8.0mA_sweepSTO2}, we display the frequency spectra measured at $I_1^*= 8$ mA while $I_2$ is changed. For these conditions, no amplitude death is observed, however a frequency branching for $I_2 \geq 8$ mA is present. We ascribe this phenomenon to the presence of an EP in the linearized model. 
Therefore, we use formula \eqref{eq:eps_condition} collocating the EP at $(I^*_{1,\text{EP}};I_{2,\text{EP}})=(8; 8)\,$mA. 
Based on this identification,  the linear coupling constant can be determined and the theory  parameters  adjusted in order to compute the eigenvalues  $i\nu_{1,2}$ as a function of the current $I_2$  (shown in fig. \ref{fig_exceptional:theory_8mA}). 
From the general point of view, the eigenvalues $i\nu_{1,2}$ give information about the linear dynamics around the rest position. 
 Their imaginary parts  show a branching similar to the experimentally observed oscillation frequencies   (fig. \ref{fig:2nd_vsSTO2_8.0mA_sweepSTO2}). 
 Indeed, the theoretical linear approach provides a good access to the analysis of the intrinsically nonlinear regime of the experimental self-oscillations. 
 In the Methods section \ref{sec_appendix:nonlinear-dynamics}, we show the consistency of the linear model with numerical computations of the mutually coupled nonlinear dynamics based on the Thiele equations. 
 Note that in a strict sense, the eigenvalues in the coupled system cannot be directly assigned to the single STVOs, but the modes must be regarded collectively. 
 However, throughout the article we label the experimental signals corresponding to the STVO in which the oscillation mode is mainly localized, as corroborated by the nonlinear simulations and the theoretical analysis of the system eigenvectors (see Methods section \ref{sec_methods:theory} \& \ref{sec_methods:nonlinear_simulations}). 
 In fig. \ref{fig_exceptional:theory_8mA}, the real part confirms that the rest-position is unstable over the entire range $I_2$, i.e. Re$(i\nu)>0$, and hence self-oscillations are stabilized. 
 The decrease of Re$(i\nu)$ in the proximity of the EP is consistent with the experimental linewidth broadening in the range $I_2 \in [7.2;8]\,$mA in fig. \ref{fig:2nd_vsSTO2_8.0mA_sweepSTO2}, where noise induced fluctuations become important due to the  vicinity of the stability axis (Re$(i\nu)=0$).

 \begin{figure*}[bht!]
\centering%
\includegraphics[width=1\textwidth]{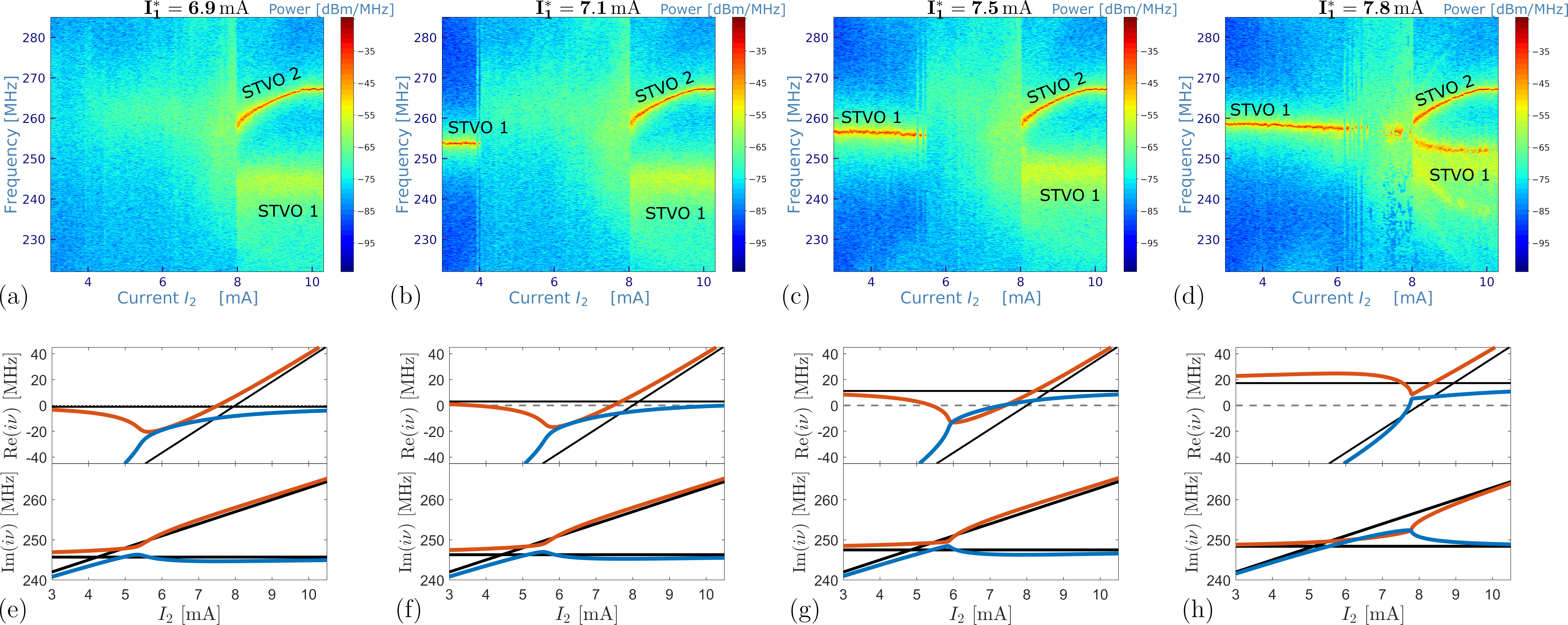}
\caption[]{ Measured frequency spectra of the coupled system vs. current $I_{2}$ in STVO 2  for different currents $I_{1}^*$ in STVO 1 (a-d). (e-h) Real and imaginary part of $i\nu_1$ (red) and $i\nu_2$ (blue) fixing current $I_1$ and changing current $I_2$. Black lines refer to the same quantity evaluated when the coupling constant is set to 0.  Experimental and theoretical graphs in the same column correspond to the same parameters. 
} 
\phantomlabel{a}{fig_exceptional:2nd_vsSTO2_6.9mA_sweepSTO2}
\phantomlabel{b}{fig_exceptional:2nd_vsSTO2_7.1mA_sweepSTO2}
\phantomlabel{c}{fig_exceptional:2nd_vsSTO2_7.5mA_sweepSTO2}
\phantomlabel{d}{fig_exceptional:2nd_vsSTO2_7.8mA_sweepSTO2}
\phantomlabel{e}{fig_exceptional:theo-1}
\phantomlabel{f}{fig_exceptional:theo-2}
\phantomlabel{g}{fig_exceptional:theo-3}
\phantomlabel{h}{fig_exceptional:theo-4}
\label{fig_exceptional:2nd_vsSTO2_sweepSTO2_diff-STO1}
\end{figure*}

From the prediction of our model, we expect that the decreasing of the gain effect (through the adjustment of the spin transfer torque in our case) together with the attraction of the eigenvalues around the EP will make the amplitude death phenomenon observable. 
 To confirm this behavior, in fig. \ref{fig_exceptional:2nd_vsSTO2_6.9mA_sweepSTO2} - \ref{fig_exceptional:2nd_vsSTO2_7.8mA_sweepSTO2},  we perform measurements of the coupled system for smaller $I_{1}^*$ for which the eigenvalue real part can explicitely become  negative due to the EP and  hence, amplitude death occurs in this regime. 
With respect to the critical currents of the uncoupled STVOs, STVO 1 is undercritical in fig.  \ref{fig_exceptional:2nd_vsSTO2_6.9mA_sweepSTO2} and overcritical in fig. \ref{fig_exceptional:2nd_vsSTO2_7.1mA_sweepSTO2} - \ref{fig_exceptional:2nd_vsSTO2_7.8mA_sweepSTO2}. 
The overall range of oscillation death evolves with $I_{1}^*$ (fig. \ref{fig_exceptional:2nd_vsSTO2_7.1mA_sweepSTO2}-\ref{fig_exceptional:2nd_vsSTO2_7.8mA_sweepSTO2}), whereas rather the smaller value $I_2$ defining the amplitude death interval is affected than the larger one which remains quasi constant. 
Increasing $I_{1}^*$ tends to stabilize the oscillation of STVO 1 and in consequence  counteracts the  occurrence of the amplitude death. 
This leads to a decrease of the current range in which no oscillation is detected (see fig. \ref{fig_exceptional:2nd_vsSTO2_sweepSTO2_diff-STO1}). 
Furthermore, for $I_1^*<7.8\,$mA and $I_2>8\,$mA, the oscillations from STVO 1 show a lower output power together with a larger linewidth than it would be expected for self-sustained oscillations. 
When the current $I_1^*$ is close to 8 mA, in the vicinity of the EP,  %
thermal noise can induce stochastic transitions between the oscillatory regime and the rest state corresponding to amplitude death (clearly visible in fig. \ref{fig_exceptional:2nd_vsSTO2_7.8mA_sweepSTO2}). 
For currents $I_{1}^* \gtrsim 8\,$mA (see fig. \ref{fig:2nd_vsSTO2_8.0mA_sweepSTO2} for $I_{1}^*=8\,$mA),  oscillation death is no more occurring, however,  the linewidth of the oscillation is clearly enhanced in a small range $I_{2}\in [7;8]\,$mA. 
This range however decreases with increasing currents $I_1^*$. 
At even larger currents $I_{1}^*\gtrsim 9\,$mA (see Methods \ref{sec_methods:experiment}), the two STVOs tend to mutually  synchronize, a phenomenon that is commonly  known for STNOs \cite{Kaka2005, Mancoff2005,Slavin2009,Lebrun2017,Tsunegi2018} and which refers to the strongly nonlinear characteristics of the oscillator, far from the Hopf bifurcation point.

The experimentally observed amplitude death  is very well reproduced by our modelling of the coupled STVOs. 
In fig. \ref{fig_exceptional:theo-1} - \ref{fig_exceptional:theo-4}, we present the corresponding real and imaginary parts of the eigenvalues $i\nu_{1,2}$ as a function of the current $I_2$. 
Except for the value of the coupling constant, which in principle depends on the electric interface between the STVOs as well as on their dynamical state, the modelling parameters are the same as those used in fig. \ref{fig_exceptional:theory_8mA}. 
We find that by only rotating the before determined coupling constant  $k_{EP}^*\rightarrow k_{EP}^*\,e^{i\phi_k}$ in the complex plane $(k_c,k_d)$ without changing its modulus,  the amplitude death phenomena can be completely described. The rotation angle for the two cases where the amplitude death is evident at $I_1^*=7.1$ and $7.5\,$mA (figs. \ref{fig_exceptional:2nd_vsSTO2_7.1mA_sweepSTO2} \& \ref{fig_exceptional:2nd_vsSTO2_7.5mA_sweepSTO2}) is $\phi_k=40$ and $45^\circ$, respectively. In fig. \ref{fig_exceptional:theo-2} - \ref{fig_exceptional:theo-3}, the amplitude death current ranges can be recognized by looking at where the condition Re$(i\nu_{1,2})\leq0$ is satisfied. Then, at the upper current value $I_2$ of the amplitude death regime, the real part of one eigenvalue crosses the real axis and the corresponding mode becomes unstable. This situation corresponds to a Hopf bifurcation which brings the system to self-oscillations. Such consideration permits to rigorously justify the presence of the upper branch in the measured spectra. 
The discussed Hopf bifurcation point does not significantly change its position while the square root like upper branch of Re$(i\nu)$ at lower currents $I_2$ implies a strong dependence of the amplitude death range's lower boundary on the fixed current $I_1^*$, as also found experimentally. 
For larger current $I_2$, in the case $I_1^* = 7.5$ mA, also the real part of the other eigenvalue (blue curve in Re$(i\nu)$) becomes positive, but it stays close to the real axis. In the experiments, which are  subject to thermal fluctuations, this manifests as the described linewidth broadening of STVO 1's oscillation at relative smaller power. 
Similar situation occurs for $I_1^* = 6.9$ mA and $I_1^* = 7.1$ mA. In both cases the  value of the rotation angle is set to $\phi_k = 40^\circ$. The main difference with the $I_1^* = 7.5$ mA case is that only the real part of one eigenvalue crosses the real axis. The other stays close to it. Similar to before, thermal fluctuations shall permit oscillations, however exhibiting a large linewidth in the experiments. 
For $I_1^* = 7.8$ mA,  the measured spectra are similar as for $I_1^* = 8$ mA and hence we set  $\phi_k=0^\circ$. 
The oscillations' death for this case is experimentally observed (see fig. \ref{fig_exceptional:2nd_vsSTO2_7.8mA_sweepSTO2}), but is not described by the linear theory. 
Note that nonlinearity might become more important in this regime. 
However, the stochasticity of the transitions between oscillation regime and rest state suggests that also thermal fluctuations play in this case a dominant role in determining the stability of the oscillators. \\
Indeed, the main characteristics of the coupled system can be accessed by the developed linearized theory. The study of the eigenvalues as a function of the  current permits to unravel the key features of the coupled STVO system's frequency response. It is important to stress that the eigenvalues refer to the eigenvectors of the matrix $A$, which are present in both STVOs. However, since the coupling is relatively small, the eigenvectors are hence mainly localized in one STVO (the labeling in figs. \ref{fig:2nd_vsSTO2_8.0mA_sweepSTO2} \& \ref{fig_exceptional:2nd_vsSTO2_6.9mA_sweepSTO2}-\ref{fig_exceptional:2nd_vsSTO2_7.8mA_sweepSTO2} refers to this aspect, see Methods section for further discussion).

In conclusion, we exploit the non-hermiticity of two coupled spintronic nano-oscillators and demonstrate the emergence of EPs in this spintronic system, which is in fact promising candidate for multiple potential applications \cite{Locatelli2013}.  
The existence of an EP drastically influences the eigenvalue characteristics leading to various complex phenomena, such as oscillation death or stochastic oscillation stability. 
We develop a theoretical modelling and show that the main experimental features at this stage can be well reproduced by linearized coupled spintronic equations.

\vspace{2ex}
\paragraph*{\bf Outlook}~

One of the interesting specificities of the spintronic nano-oscillators is their strong nonlinearity which makes them a promising candidate for various applications and leads to a tremendous manifold of physical phenomena unified in these nanoscale devices. 
The emergence of an EP in a nanoscale nonlinear system is to our opinion of fundamental interest. 
Beyond the already mentioned implications for the development of novel types of sensors operating at exceptional points \cite{Wiersig2014,Chen2017}, these systems are anticipated to unravel fascinating physics. 
This includes phenomena such as chaos, complex bifurcations, or the emergence of topological operations around the EP \cite{Xu2016,Roehm2018}. 
Complex dynamics and as well the demonstrated occurrence of stochastic stability might furthermore  complement the field of hardware based neuromorphic computing that recently gained attention in the context of spintronics\cite{Grollier2016}, for instance as stochastic spiking neurons. 
Non-hermiticity in this respect adds an additional complex response of the system to input signals\cite{Amir2016,Tanaka2019}, implying abrupt phase transitions which are also inherent in neural networks\cite{Tognoli2014}. Characteristics of non-hermiticity have been found in the description of the brain, for instance in EEG measurements \cite{Marzetti2008,Tozzi2017}, or the inhibitory and excitatory balance in neocortical neurons \cite{Haider2006}, similar to nonconservative elements of gain and loss in our STNO system. 
We emphasize that higher dimensionally coupled systems have been realized with STNOs\cite{Tsunegi2018,Zahedinejad2019} which are anticipated to  facilitate the emergence of higher order exceptional points \cite{Hodaei2017, Wang2021,Yu2020} or other complex dynamics \cite{Flebus2020,Gunnink2022}. 
All these different aspects  are still to be explored and  potentially lead to intriguing findings in nonlinear  non-Hermitian systems.

\vspace{4ex}

\renewcommand*{\bibfont}{\scriptsize}

\newlength{\bibitemsep}\setlength{\bibitemsep}{1pt}
\newlength{\bibparskip}\setlength{\bibparskip}{0pt}
\let\oldthebibliography\thebibliography
\renewcommand\thebibliography[1]{%
  \oldthebibliography{#1}%
  \setlength{\parskip}{\bibitemsep}%
  \setlength{\itemsep}{\bibparskip}%
}

\bibliographystyle{naturemag}
\bibliography{literatur_promo}

\clearpage

\paragraph*{\bf \large Methods}~ 
\vspace{3ex}

\small
\setcounter{figure}{0}
\renewcommand{\thefigure}{S\arabic{figure}}
\captionsetup{font=footnotesize}

\paragraph{\bf Theory} ~ \vspace{1ex}
\label{sec_methods:theory}

\paragraph*{\bf \small Thiele's equations.}
\label{Sec:Thiele's theory}

The model equations for the two coupled 
STVOs are the following \cite{Bortolotti2013}:
\begin{align}
\frac{d\boldsymbol{\rho}_i}{dt} =& \Omega_i(\rho_i,I_i) \boldsymbol{n}_i \times \boldsymbol{\rho}_i + \left[  C_i I_i - d_i\cdot \Omega_i(\rho_i,I_i)\right] \boldsymbol{\rho}_i     \nonumber \\ 
&+ K \cdot \boldsymbol{\rho_j}\, , %
\label{eq:coupled-DGL}
\end{align}
where $i,j \in \{1,2\}, i\neq j$, denoting STVO 1 or 2. The symbol $\boldsymbol{n}_{i}$ denotes the unit vector along the symmetry axis of the oscillator, $I_i$ is the injected current,  the variable $\boldsymbol{\rho}_i $ is the  in-plane vortex core position normalized to the radius $R_i$ of the device, $\Omega_i$ is the conservative oscillations angular frequency,  $d_i$ a dimensionless damping constant and $C_i$ is a parameter determining the efficiency of the 
spin torque effect. This latter parameter determines also the critical current $I_{c,i}$ needed for exciting self-sustained oscillations in the uncoupled case.    
The angular frequency $\Omega_i$ is a function\ of $I_i$ and of the magnitude $\rho_i = | \boldsymbol{\rho}_i |$:
$\Omega(\boldsymbol{\rho},I) = \Omega_{ms} (1/[1-(\rho/2)^2]) + n_{oe} I (1- \rho^2/2)$, where $\Omega_{ms}$ and $n_{oe}$ take into account 
the influence on the frequency of the magnetostatic field generated by the magnetic vortex
state and of the Oersted field generated by the injected current, respectively. 
The coupling between STVO $i$ and $j$ is described by a $2\times 2$ matrix:
\begin{align*}
K\cdot \boldsymbol{\rho}_j = k_d \boldsymbol{\rho}_j + k_c \boldsymbol{n}_j \times \boldsymbol{\rho}_j  .
\end{align*}
where $k_c$ and $k_d$ are conservative and dissipative coupling coefficients, respectively.

\vspace{2ex}
\paragraph*{\bf \small $\mathcal{PT}$-symmetry.}
\label{sec_methods:PT-symmetry}

It is interesting to notice that at perfect gain compensation, $\beta_1 = -\beta_2=\beta$,
and the same frequencies of the uncoupled oscillators, $\omega_1 = \omega_2 = \omega$, the natural frequencies
are given by the simplified formula: $\nu_{1,2} = {\omega} \pm \sqrt{ k^2  - \beta^2}$.
Thus, the natural frequencies are both real when $\beta < k$, while for $\beta >k$ an imaginary
part appears. The system has in this case a $\mathcal{PT}$-symmetry when the frequencies are real 
and a broken $\mathcal{PT}$-symmetry when the frequencies are complex. The symmetry-breaking bifurcation
occurs when $\beta = k$ and this condition corresponds to an EP. Representing an interesting case, this situation however requires
a very fine tuning of the two oscillators' properties.

\vspace{2ex}
\paragraph*{\bf \small System parameters.}
\label{sec_methods:parameters}

\begin{table}[bth!]%
 \caption{\label{tab_methods:parameters} System parameters for the calculated eigenvalues. }
 \begin{ruledtabular}
 \begin{tabular}{cc}
 $\Omega_{ms,1}\,$[MHz] & $2\pi\cdot 225$  \\
 $\Omega_{ms,1}\,$[MHz]   &  $2\pi\cdot 233$   \\
$n_{Oe,1}=n_{Oe,2}\,$[MHz/mA]  & $2\pi\cdot 3$  \\
$d_1=d_2$  & $0.1$  \\
$C_1\,$[MHz/mA] & $22.22$  \\
$C_2\,$[MHz/mA] & $20.18$ 
 \end{tabular}
 \end{ruledtabular}
 \end{table}

The used model parameters of the system, derived from the experimental data, are listed in table \ref{tab_methods:parameters}.

\vspace{2ex}
\paragraph*{\bf \small Exchange of eigenvectors.}
\label{sec_methods:eigenvectors}

The oscillations in practice take place in the two oscillators. However, in the coupled system the dynamics must be rather understood as a collective phenomenon. In this respect, it is interesting to correlate the eigenvectors with their localization in each STVO. 

\begin{figure}[bth!]
 \centering  %
   \includegraphics[height=0.502\textwidth]{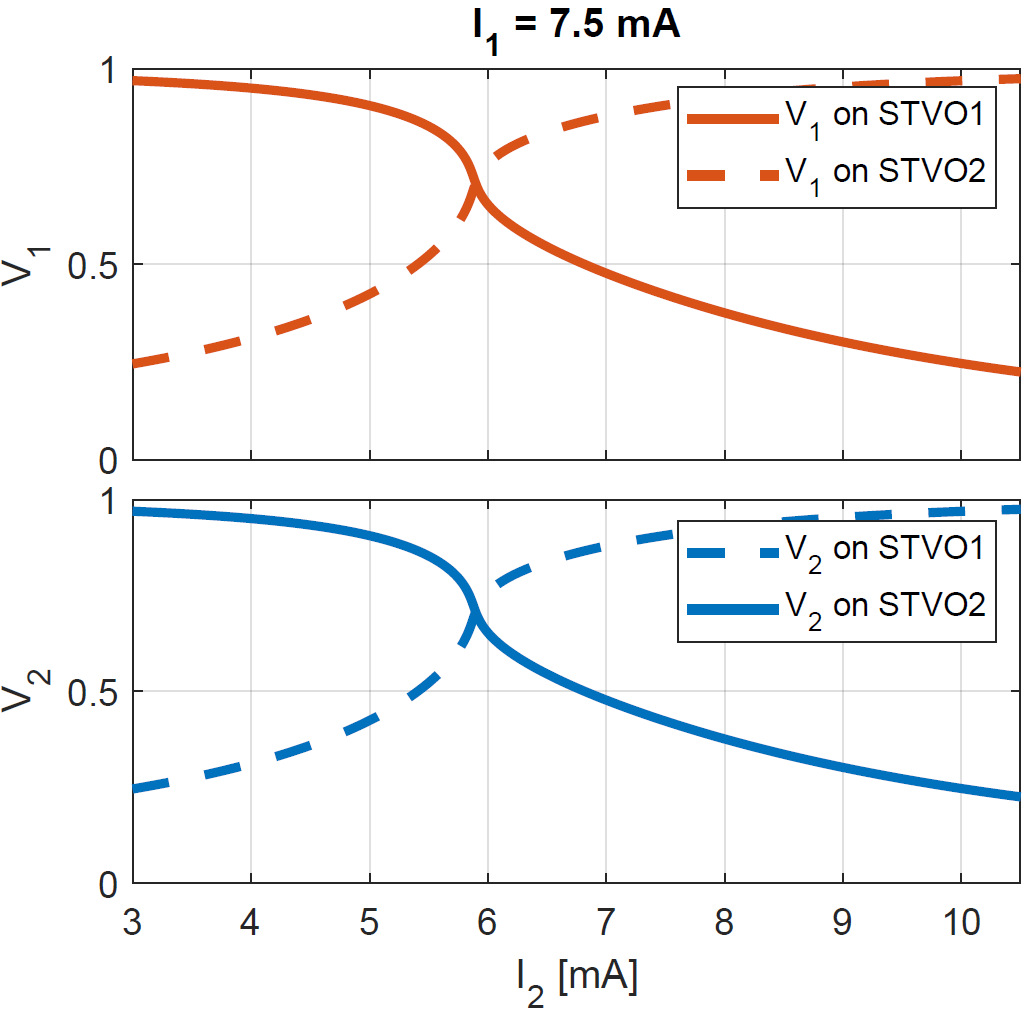} %
   \caption[]{Projection of the system's eigenvectors onto the STVOs in order to determine their localization. }
  \label{fig_methods:eigenvectors}
\end{figure}

In fig. \ref{fig_methods:eigenvectors}, we plot the projection of the two eigenvectors $v_1$ and $v_2$ on STVO 1 and STVO 2 at an exemplary current value of $I_1^* = 7.5\,$mA (see also fig. \ref{fig_exceptional:theo-3}).  
When the eigenvalues approach each other (fig. \ref{fig_exceptional:theo-3}), there is an exchange of the  eigenvectors' components. 
In fig. \ref{fig_methods:eigenvectors}, the eigenvector $v_1$ is localized mainly in STVO 1 and $v_2$ in STVO 2 for current values of $I_2\lesssim 6\,$mA. For $I_2\gtrsim 6\,$mA, the eigenvectors change their localization and $v_1$ ($v_2$) is majorly identified in STVO 2 (1).

\vspace{3ex}
\paragraph{\bf Nonlinear dynamics} ~ \vspace{1ex}
\label{sec_appendix:nonlinear-dynamics}

The results presented in section Experiments have their theoretical foundation in the linearized theory of the coupled system dynamics although the measured spectra refer to phenomena (e.g self-oscillation regimes) which can only be rigorously explained with nonlinear theory arguments. It is therefore important to show the consistency of the linear model with computations of nonlinear dynamics of the coupled STVOs.

\vspace{2ex}
\paragraph*{\bf \small Numerical model.}
\label{sec_methods:numerical_model}

\begin{figure}[bth!]
  \centering  %
  \subfloat[  \label{fig_methods:STO1_alone}] 
  {  \raisebox{0.0cm}{ \hspace{0.0cm} 
  \includegraphics[height=0.38\textwidth]{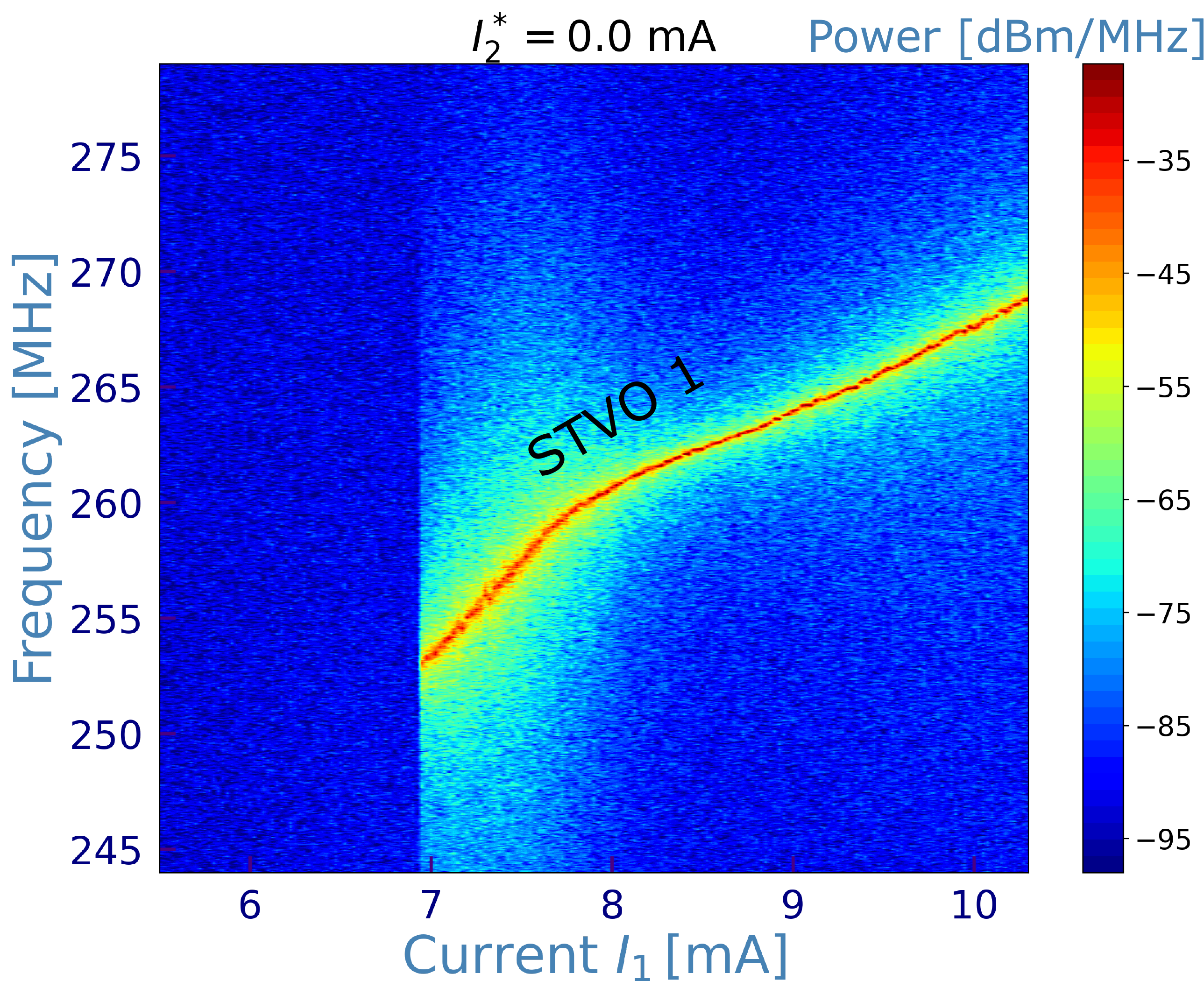} } 
  }
  \subfloat[  \label{fig_methods:STO2_alone}]
  {  %
  \includegraphics[height=0.38\textwidth]{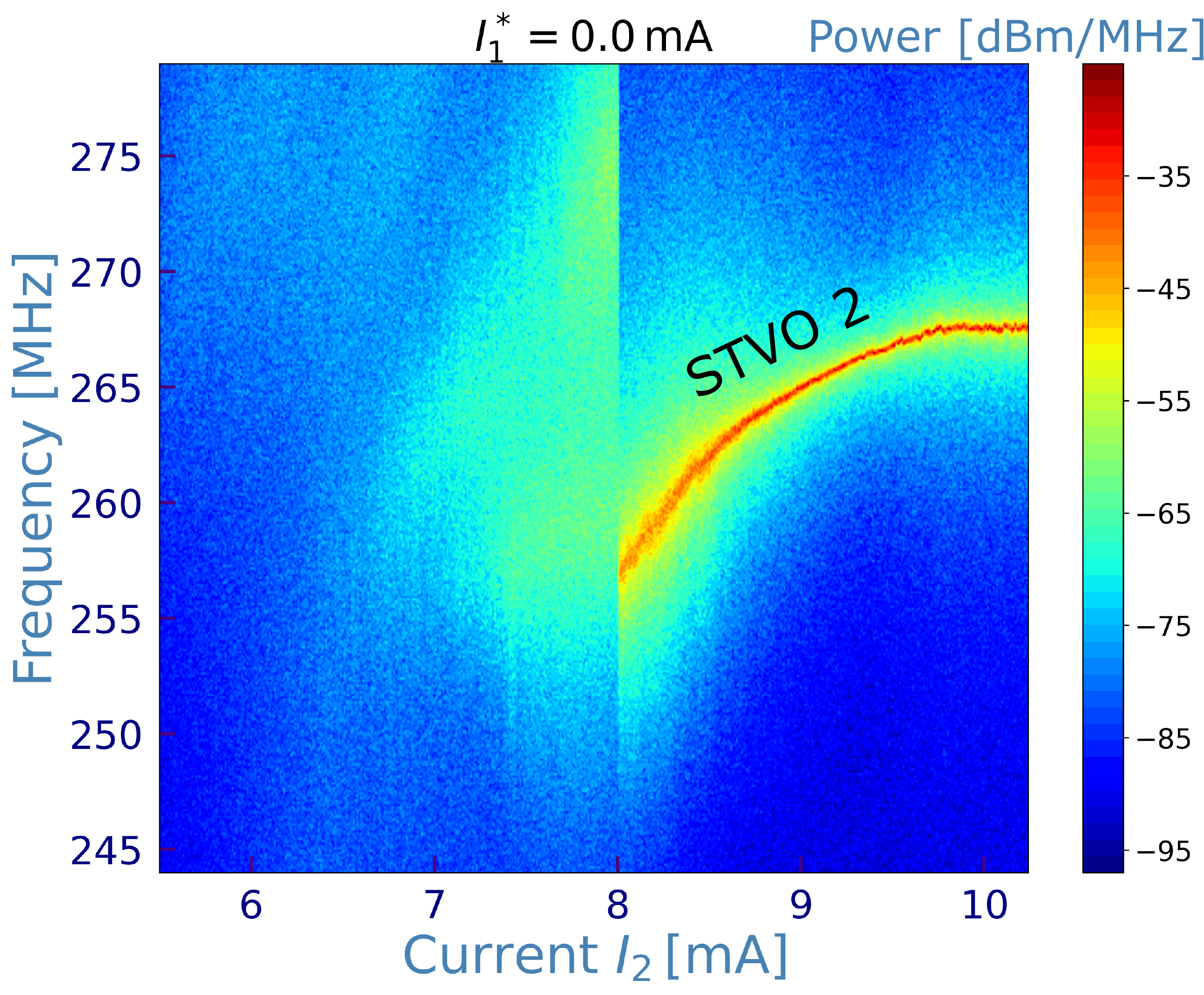}  }   
   \caption[uncoupled]{Frequency spectra of the two independent STVOs. 
}    
  \label{fig:uncoupled-spectra}
\end{figure}

Based on the single STVO measured spectra (shown in fig. \ref{fig:uncoupled-spectra} further below), the parameters of the Thiele equation for each oscillator have been determined. In particular, from the critical current $I_{c,i}$ and the corresponding self-oscillation frequency, the value of the constants $\Omega_{ms,i},\,C_i$ assuming certain values for $n_{Oe}$ and $d_i$ can be estimated. The nonlinear current dependence of the measured self-oscillation frequency is reproduced by assuming a polynomial dependence on the vortex core position of the damping term $d_i\rightarrow d_i(\rho_i)$. We remark that this modeling procedure is for the single STVO uncoupled from  the other one. The coupling effect is determined as described in the section Experiments.

\vspace{2ex}
\paragraph*{\bf \small Simulations.}
\label{sec_methods:nonlinear_simulations}

\begin{figure}[bth!]
  \centering  %
  \subfloat[  \label{fig_methods:theo1}] 
  {  \raisebox{0.0cm}{ \hspace{0.0cm} 
  \includegraphics[height=0.38\textwidth]{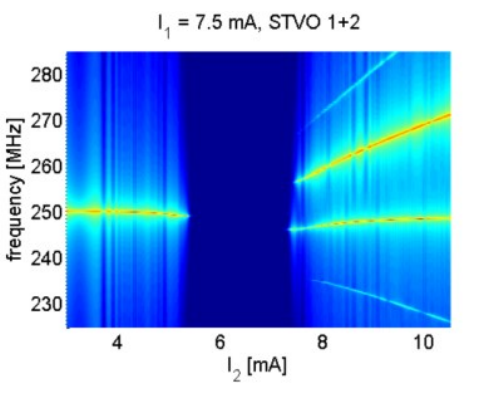} } 
  }
  \subfloat[  \label{fig_methods:theo2}]
  {  %
  \includegraphics[height=0.38\textwidth]{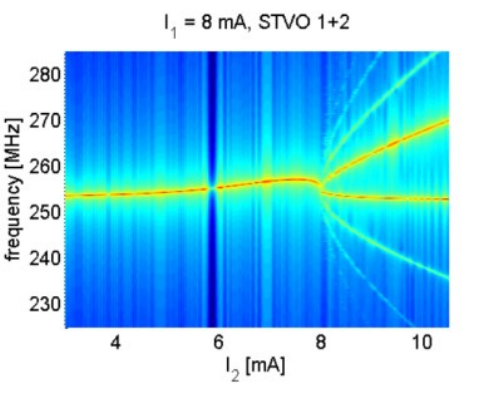}  }   
   \caption[uncoupled]{Frequency spectra computed by integrating the nonlinear model of the coupled STVOs for two current $I_1^*$ values. The color scale is in arbitrary units and is analogous to figures which show the measured spectra. }
  
  \label{fig:nonlinear-model-spectra}
\end{figure}

In fig. \ref{fig:nonlinear-model-spectra}, the computed spectra  from time integration of the nonlinear dynamics of the coupled STVOs are shown. 
Fixing the current $I_1$, the dynamics of the coupled oscillators is simulated for each value of the current $I_2$ in the range $(3,10.5)$ mA, taking an ensemble of $N$ initial conditions randomly distributed in a disk around the orgin of radius $\rho/R=0.01$. Then from the self-oscillations regime we estimate the power as: $P = 1/N\,\log_{10}(\sum_i |\tilde{x}_{1,i}+\tilde{x}_{2,i}|^2)$, where $\tilde{x} = FFT[x]$, and the notation $x_{n,i}$ refers to the x coordinate of the vortex core of the STVO-n.

The spectra computed from the nonlinear model confirm the validity of the approach used in the Experiments section  to estimate the coupling constant $k_{EP}$. The similarity of fig.  \ref{fig:2nd_vsSTO2_8.0mA_sweepSTO2} and  fig. \ref{fig_methods:theo2} is a strong indication of the fact that the branching of the frequency in the measured spectra is due to the passage in the proximity of an EP. The agreement between experimental results and numerical simulations in terms of frequency value and frequency gap when the spectra become double peaked is obtained also for the other cases where the current $I_1^*$ is changed in the range $[6.9 ; 8]$ mA. Finally, the agreement on the amplitude death current ranges is not surprising since the use of the linear theory for their estimation is a rigorous result of the nonlinear system theory.

\vspace{3ex}
\paragraph{\bf Experimental description} ~ \vspace{1ex}
\label{sec_methods:experiment}

\paragraph*{\bf \small Device fabrication.}
\label{sec_methods:device-fabrication}

The studied STVO devices are magnetic tunnel junctions containing a pinned layer made of a conventional synthetic antiferromagnetic stack (SAF), a MgO tunnel barrier and a  NiFe free layer in a magnetic vortex configuration (blue, green and yellow layers in fig. \ref{fig:circuit}, resp.).
The  magnetoresistive ratio related to the tunnel magnetoresistance effect (TMR) lies around $110\,$\% at room temperature and the area resistance product is $RA \approx 2\,\Omega\upmu\text{m}^2$.  %
In detail, the SAF is composed of \selectlanguage{ngerman} IrMn($60$)/""Co$_{70}$Fe$_{30}$($2.6$)/""Ru($0.85$)/""Co$_{40}$Fe$_{40}$B$_{20}$($2.6$) and the total layer stack is Ta(5)/""CuN(50)/""Ta(5)/""Ru(5)/""SAF/""MgO($1$)/""Co$_{40}$Fe$_{40}$B$_{20}$($2$)/""Ta($0.2$)/""Ni$_{80}$Fe$_{20}$(7)/""Ta(10)/CuN(30)/""Ru($5$), with the nanometer layer thickness in brackets.
\selectlanguage{english}
The growth of the amorphous NiFe free layer is decoupled from the lower CoFeB layer by a $0.2\,$nm Ta-layer. This structure permits to exploit the high tunnel magnetoresistance (TMR) ratio of the crystalline CoFeB-junction and the magnetically softer NiFe for the vortex dynamics.   
The layers are deposited on high resistivity SiO$_2$ substrates by ultrahigh vacuum magnetron sputtering and subsequently annealed for $2\,$h at $T=330\,^{\circ}$C at an applied magnetic field of $1\,$T along the SAF's easy axis. 
The patterning of the circular tunnel junctions is conducted using e-beam lithography and Ar ion etching. 
They have an actual diameter of $2R = 370\,$nm and the microwave field line of $1\,\upmu\text{m} \times 300\,$nm is lithographied $300\,$nm above the nanopillar.

\vspace{2ex}
\paragraph*{\bf \small Working principle and measurements.}
\label{sec_methods:measurements}

Measurements are conducted under an applied out-of-plane field of $\mu_0 H_{\perp} = 360\,$mT. It tilts the in-plane magnetization of the SAF slightly into the perpendicular direction. That induces an out-of-plane spin current polarization, necessary for an efficient spin transfer torque (STT) \cite{Dussaux2010}.  
The STT provides the gain mechanism in the STVOs and is generated by  injected dc currents, which are separately controlled for the two STVOs by two dc current sources. 
The TMR effect converts the vortex magnetization dynamics into an electrical rf signal. 
The electrical rf output signal of each oscillator is amplified by $30\,$dB and injected into the field line located above the other STVO (see fig. \ref{fig:circuit}) in order to implement a symmetric coupling scheme through the generated rf Oersted fields. 
In the electrical circuit, the dc and rf current parts are separated through a bias tee and the dc electrical properties are monitored by a voltmeter. 
 The emitted rf  signals of both coupled STVOs are combined and recorded by a spectrum analyzer.

\vspace{2ex}
\paragraph*{\bf \small Symmetry of the coupled STVO system.}
\label{sec_methods:uncoupled-characteristics}

In order to reveal the emergence of an exceptional point and  complex dynamics in the coupled STVO system, the uncoupled characteristics of the oscillators should be sufficiently similar for approximately  realizing a symmetric situation with reciprocal coupling.  

In fig. \ref{fig:uncoupled-spectra}, the power spectra  for the two uncoupled STVOs are measured separately. They show that in the vicinity of the threshold current for self-oscillations the frequency characteristics of the two STVOs are very similar.  The region close to the onset of oscillations is the one of interest for the study of EPs.  

\begin{figure}[bth!]
\centering  %
\includegraphics[height=0.72\textwidth]{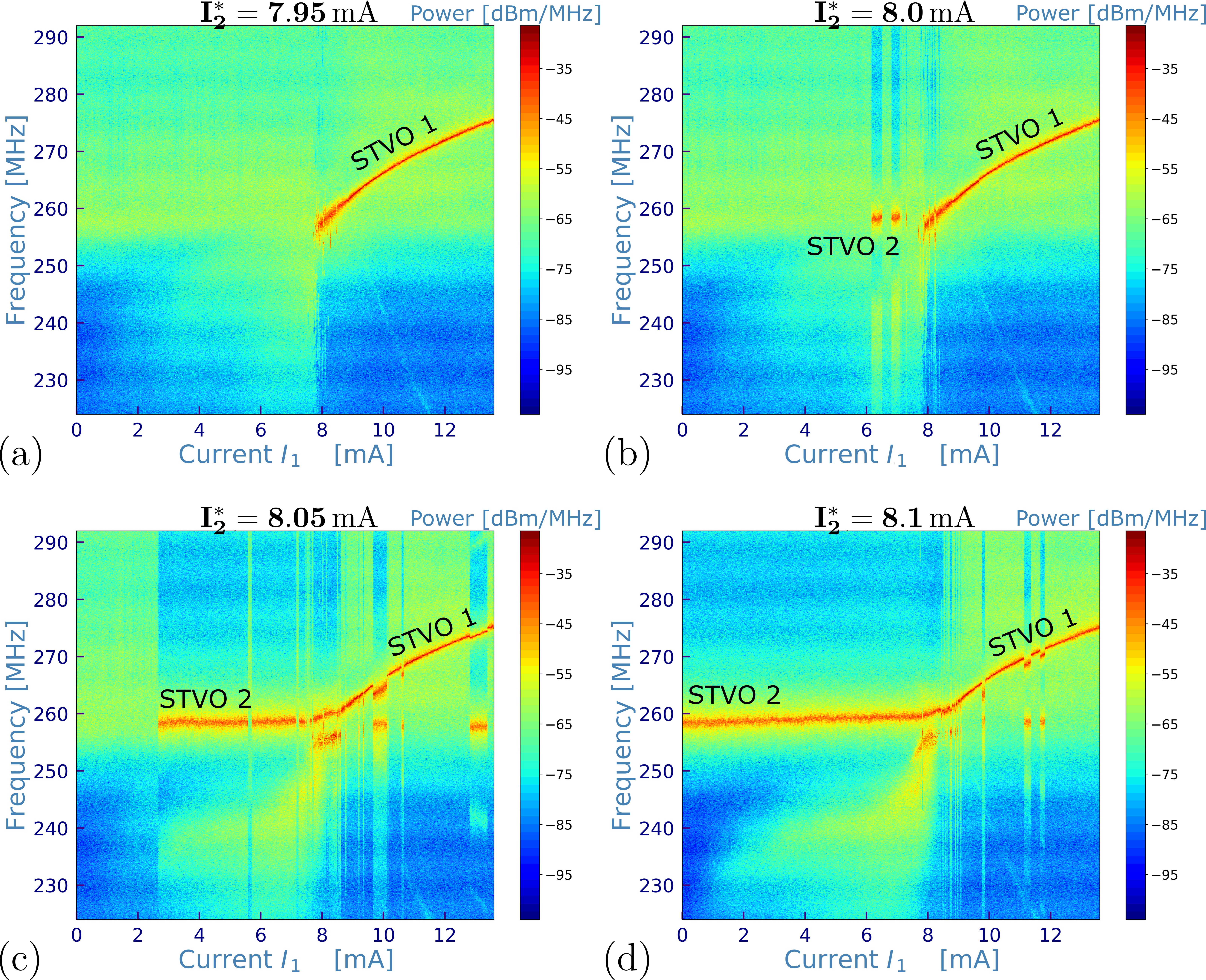} %
\protect\caption{Measured frequency spectra of the coupled system vs. current $I_1$ for different constant currents $I_2^*$. }
\label{fig_methods:vs_I1}
\end{figure}

However, the realization of a perfectly symmetric system is experimentally difficult owing to small deviations in the complex, nanometer scale fabrication process, and the coupled system is very sensitive to small parameter changes in the vicinity of the EP. 
Performing measurements at constant current $I_2^*$ while changing the current $I_1$ in STVO 1 (fig. \ref{fig_methods:vs_I1}) provides qualitatively different results than the ones shown beforehand fixing the current in STVO 1. 
We observe a non-trivial behavior of the coupled system with either suppressed or stabilized oscillations of the STVOs. At $I_2^*=7.95\,$mA (fig. \ref{fig_methods:vs_I1}a), STVO 2 is not sustained, but the critical current of STVO 1 slightly shifted to a larger current (see fig. \ref{fig_methods:STO1_alone}). 
At $I_2^*=8\,$mA (fig. \ref{fig_methods:vs_I1}b), corresponding to the critical current of STVO 2 (see fig. \ref{fig_methods:STO2_alone}), the latter's auto-oscillations are suppressed for $I_1 \lesssim 6\,$mA, and are subject to stochastic effects above $6\,$mA until the oscillations of STVO 1 become stabilized. 
For  $I_{2}^*=8.05\,$mA and  $I_{2}^*=8.1\,$mA (figs. \ref{fig_methods:vs_I1}c \& \ref{fig_methods:vs_I1}d, resp.), we observe the stabilization of STVO 2's auto-oscillations at smaller currents $I_1$ and  a suppression of STVO 2's oscillations after its frequency has crossed that of STVO 1 at $I_{1} \gtrsim  8.5\,$mA. 

\begin{figure}[bth!]
 \centering  %
  \includegraphics[width=0.465\textwidth]{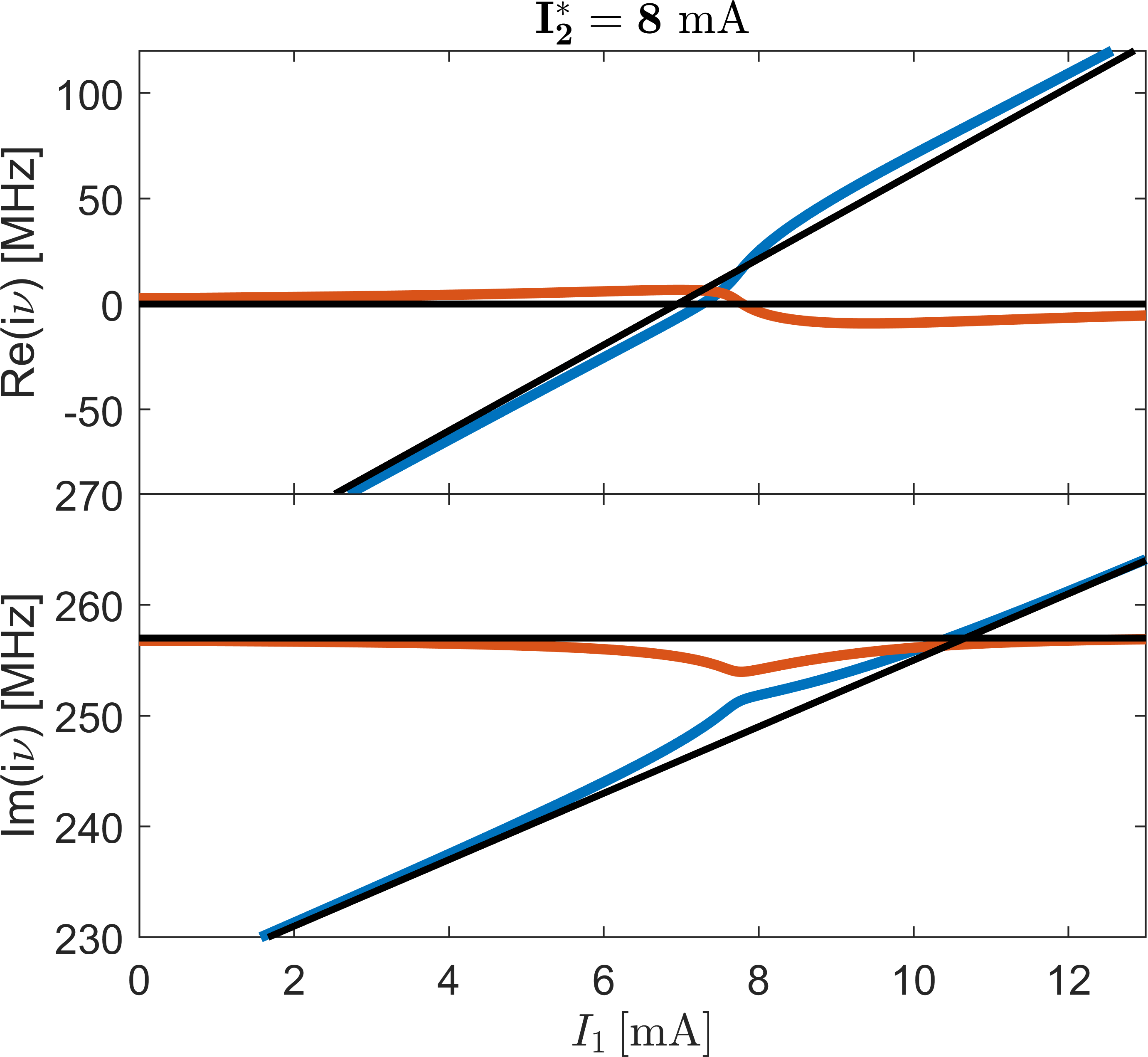} %
  \caption{Theoretically modeled eigenvalues vs. current $I_1$ for  constant current $I_2^* = 8\,$mA. All parameters are chosen identical as the ones at constant currents $I_1^*$ shown above. The coupling phase is set to $\phi_k=10\,^{\circ}$. }
  \label{fig_methods:inverse_theo}
\end{figure}

In fig. \ref{fig_methods:inverse_theo}, we show that also here the linearized theory qualitatively reproduces the observed features, employing the same parameters as for the inverse case presented in the main text, and a coupling phase of $\phi_k=10^{\circ}$. Note that in this case, an extreme fine tuning of the theoretically applied parameters would be necessary in order to thoroughly reproduce the experimental characteristics since already a very subtle change of $I_2^*$ in the vicinity of the EP significantly changes the oscillation properties. Hence, here we choose $I_2^*=8\,$mA as a representative current value. 
It can be observed that Re$(i\nu_2)$ (red-curve)  is close to the zero stability axis. Increasing $I_1$ from 0 first increases Re$(i\nu_2)$ leading to possible stabilized oscillations up to $8\,$mA, such as reflected experimentally in figs. \ref{fig_methods:vs_I1}b - \ref{fig_methods:vs_I1}d. A larger current $I_2^*$ stabilizes the oscillations  already at lower $I_1$ while a smaller one leads to a shift of  Re$(i\nu_2)$ entirely below the zero axis, also reflected experimentally in fig. \ref{fig_methods:vs_I1}. 
For current values of $I_1 \gtrsim 8\,$mA, Re$(i\nu_2)$ (red-curve) becomes negative while staying close to the stability axis. This leads to oscillation death in this regime while partly seeing oscillations when thermally induced transitions into the positive plane occur (see  fig.\ref{fig_methods:vs_I1}c - \ref{fig_methods:vs_I1}d). 
The value $I_1$ at which the blue curve in fig.  \ref{fig_methods:inverse_theo} crosses the zero axis is slightly larger than at zero coupling (black curve), reproducing the observed larger critical current at $I_2^*=7.95\,$mA in fig. \ref{fig_methods:vs_I1}a. 

The measurements emphasize the physical richness of the coupled system even at only little asymmetry between the two oscillators. 
It can be stated that the individual parameters of the two STVOs are of critical importance, especially when being close to an EP. Furthermore, the nature of the coupling between the two oscillators is responsible for the observed characteristics and leads to manifold interesting effects in the collective system and hence, will be an additional important control parameter in future experiments.

\paragraph*{\bf \small Synchronization at larger current densities.}
\label{sec_methods:synchro}

\begin{figure}[bth!]
\centering  
\includegraphics[width=0.96\textwidth]{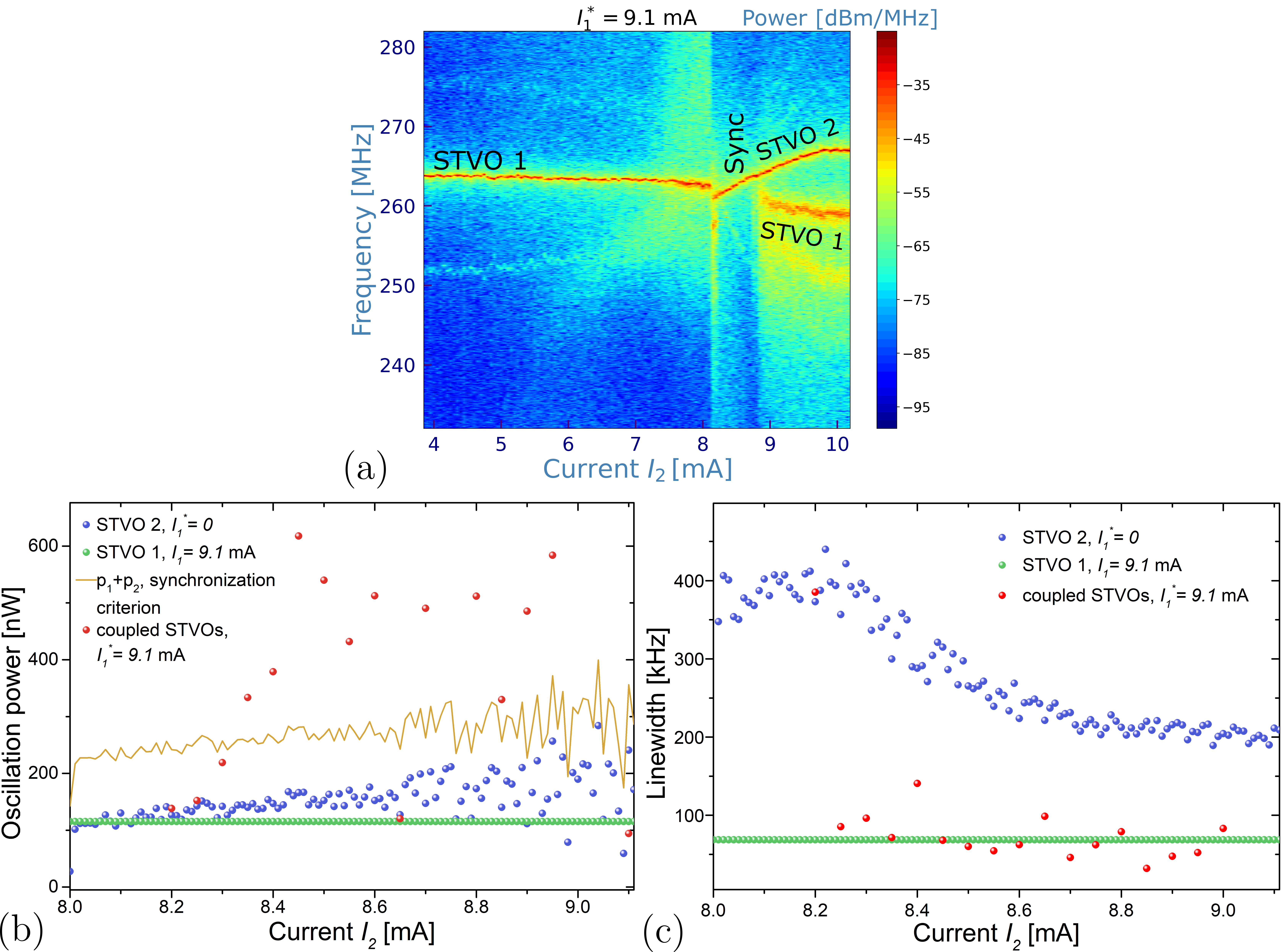}%
\caption{Experimental characterization in the regime of mutual synchronization of the two coupled STVOs: $I_1^*=9.1\,$mA, $\mu_0 H_{\perp} = 360\,$mT, $T = 300\,$K. (a) Measured spectra. (b-c) Oscillation power and linewidth  evaluated for the synchronized and the uncoupled case. }
\label{fig_methods:synchro}
\end{figure}

In fig. \ref{fig_methods:synchro}a, we show the power spectra recorded for larger current densities $I_1^*$ in our experiment. We observe the well-known, intrinsically nonlinear effect of mutual synchronization  between $I_2 = 8.2$ and $8.95\,$mA. Such synchronization implies an increase of the emitted power and a decrease of the spectral linewidth as shown in figs. \ref{fig_methods:synchro}b-c. The dark-yellow line in fig. \ref{fig_methods:synchro}b defines the mutual synchronization criterion \cite{Slavin2009} beyond which the synchronized power indicates efficient synchronization. 
Note that the mutual synchronization phenomenon is the best-explored in the nonlinear coupled dynamics of STNOs and it is noteworthy that here, it co-exists with the described complex dynamics at lower current densities linked to non-hermitian physics, with the only practical difference in the supplying current. However, we want to emphasize that mutual synchronization in this context must not be necessarily taken for granted, but, depending on the system parameters, such as especially the type and strength of the coupling, different complex situations might occur (chaos, bistability, etc.), and need to be explored.

\scriptsize

\vspace{4ex}
\paragraph*{\bf\small Data Availability} ~ \vspace{1ex}

The data that support the findings of this study are available from the corresponding author upon reasonable request.

\vspace{4ex}
\paragraph*{\bf\small Acknowledgements} ~ \vspace{1ex}

S.W. acknowledges financial support from Labex FIRST-TF under contract number ANR-10-LABX-48-01. 
The work is supported by the French ANR project ''SPINNET'' ANR-18-CE24-0012 and the Horizon2020 Framework
Program of the European Commission, under FETProactive Grant agreement No.899646 (k-NET)

\vspace{4ex}
\paragraph*{\bf\small Author contributions} ~ \vspace{1ex}

S.W., R.L., C.S. and V.C. conceived the project. R.F. and R.D. prepared the devices. 
S.W. performed the experimental measurements and analyzed the data with the help of V.C., R.L. and P.B. 
C.S. and S.P. developed the theoretical model with the help of S.W.  S.P. conducted the numerical simulations. 
S.W., V.C., R.L., C.S. and S.P. prepared the manuscript and all authors discussed and contributed to the final version. 

\vspace{4ex}
\paragraph*{\bf\small Competing Interests} ~ \vspace{1ex}

The authors declare no competing Interests.

\end{document}